\documentstyle[aps,prl,twocolumn]{revtex}

\begin{document}
\draft
\title{ Fermi-liquid features of the one-dimensional Luttinger liquid}
\author{Yupeng Wang$^{1,2}$}
\address{$^1$Laboratory of Ultra-Low Temperature Physics, Chinese Academy of
Sciences, P.O. Box 2711, Beijing 100080, China\\
$^2$Institut f\"{u}r Physik, Universit\"at Augsburg, 86135 Augusburg,
Germany}
\maketitle
\begin{abstract}
We show that the one-dimensional (1D) electron systems can also be described
 by Landau's phenomenological Fermi-liquid theory. Most of the known results derived from
 the Luttinger-liquid theory can be retrieved from the 1D Fermi-liquid theory.
 Exact correspondence between the Landau 
parameters and Haldane parameters is established. The exponents of the dynamical correlation functions and the impurity
problem are also discussed based on the finite size corrections of elementary excitations with the
predictions of the conformal field theory, which provides a bridge between the 1D Fermi-liquid 
and the Luttinger liquid.
\end{abstract}
 
Landau's Fermi-liquid theory (FLT)\cite{1} is a fundamental paradigm in condensed 
matter physics.  However, as suggested by many 
authors, the situation in reduced dimensions may be different. Especially in
 1D, it is generally believed that the systems are marginal\cite{2,3,4} and the  low 
temperature properties of the 1D gapless quantum systems can be described by
 the Luttinger-liquid theory (LLT)\cite{5,6}. Even in 2D, Anderson showed that a 
strongly correlated electron system may exhibit the main characters of a 
Luttinger liquid\cite{7}.  A Luttinger liquid generally has the following 
characteristics: (i). Its low energy excitations can be completely described
 by the collective density waves, i.e., the sound
 waves (charge density fluctuation) and the spin density waves  
characterized by two velocities $v_\rho$ and $v_\sigma$ respectively; (ii). 
Various operators exhibit anomalous dimensions determined by two stiffness 
constants $K_\rho$ and $K_\sigma$, which lead to the nonuniversal power-law 
decay of correlation functions; (iii). The momentum distribution of the 
physical particles is continuous at the Fermi surface.
 
\par
In fact, the low temperature properties of a 1D gapless electron system are
 uniquely determined by four constants $v_\rho$, $v_\sigma$, $K_\rho$ and 
$K_\sigma$ and the Haldane relations hold:
\begin{eqnarray}
v_{N_\gamma}v_{J_\gamma}={v_\gamma}^2, {~~~~~}v_{J_\gamma}=K_\gamma v_\gamma,
 {~~~~}\gamma=\rho, \sigma{~~~},
\end{eqnarray}
where $v_{N_\gamma}$ and $v_{J_\gamma}$  ($\gamma=\rho,\sigma$) are the 
velocities of excitations induced by additional charges and currents, 
respectively\cite{5,6}. Some important thermodynamic quantities such as the 
specific heat $C$, the susceptibility $\chi$ and the compressibility $\kappa$
 are given by
\begin{eqnarray}
C=\frac \pi 3T({v_\rho}^{-1}+{v_\sigma}^{-1}),{~~~}\chi=\frac 1{\pi 
v_{N_\sigma}},{~~~}\kappa=\frac 2{\pi v_{N_\rho}n^2},
\end{eqnarray}
where $n$ is the density of electrons. Notice here we have taken the Boltzman
 constant $k_B$, the Bohr magneton $\mu_B$ and the Planck constant $\hbar$ as
 our units.
\par
A notable fact is that the thermodynamics of a 1D quantum system remains
that of a Fermi liquid, while the transport and dynamical properties do not.
We note that the thermodynamics depends strongly on the statistics (with or without the Pauli's exclusion)
rather than on the dimension of the system, while the transport
 coefficients are tensors and must be dimension-dependent. Therefore, a question arises:
 Does Landau's FLT completely break down in 1D or in some situations,
 it is still applicable?
 \par
 People against FLT in 1D is mainly due to the following reasons (see Ref.6
 and the references therein): (i).There is generally spin-charge separation in 1D;
 (ii).The quasiparticle residual $Z_{p_F}$ at the Fermi surface is zero due to the
 orthogonal catastrophe, and the relaxation rate of quasiparticles has the same order
 of the quasiparticle energy even near the Fermi surface, which makes the quasiparticle
 may not be well-defined. We start our discussion by checking these differences between the 3D 
Fermi liquid and the 1D quantum 
 systems. As well known, spin-charge separation effect also exists in the collective modes 
 of a Fermi liquid. For example, the velocity of charge density fluctuation generally has a 
different
 value from that of the spin fluctuations. The two sets of Landau parameters
 $F^s_l$ and $F_l^a$ are responsible to the charge and spin phenomena respectively.  In fact, spin-charge
 separation is only a collective effect rather than a single-particle effect. Therefore,
 it may not be an obstacle to apply Fermi-liquid theory in 1D. The most serious problem in 1D 
is
 the zero quasiparticle residual and the large scattering rate of quasiparticles.
  However, we note that the momentum
 distribution of the quasiparticles $<n_{ps}>$ defined in the microscopic
  theory is rather different from the quasiparticle number $n_{ps}$ 
defined by Landau. The former is only the mean value of the real particle 
number in the ground state and has no cusp at the Fermi surface for a 1D 
system, while the latter indeed has a perfect jump at the Fermi surface. 
In Landau's theory,
$n_{ps}$ is nothing but the perfect Fermi-Dirac distribution function with a
 $n_{ps}$-dependent quasiparticle energy $\epsilon_{p}$. 
Recently, some authors demonstrated that the Luttinger theorem is also 
correct for the 1D systems\cite{8}. That means some characters of the non-interacting
system must remain in the corresponding interacting system. Traditionally, the 
quasiparticle life-time are defined either by the imaginary part of the self energy
or by the relaxation time in the semi-classical theory.
 For the first case, we note that if we use the Green's function
of the quasiparticles rather than that of the real particles, the imaginary part
of the self energy may be very small compared to the real part, as long as the creation and 
annihilation
operators of the quasiparticles
are well-defined. This can be seen in the following discussion for the integrable models.
For the second case, we note that the relaxation time is nothing but the response time
of a quasiparticle to an external perturbation. Without external perturbation,
the system must be in equilibrium  and the collision integral is exactly
zero\cite{9}. Therefore, the quasiparticle relaxation time makes no sense to the equilibrium
properties. To clarify this, we consider the pure states (eigenstates) of a 1D system. For a 
finite 
non-interacting 1D system, the quantum states are described by a sequence of integers $I_j$.
Each $I_j$ corresponds to a single particle state with the momentum $2\pi I_j/L$. If we turn on
the interaction continuously, the distribution $\{I_j\}$  must not 
be changed since
$I_j$ are discrete while the interaction is continuous. Therefore, an eigenstate of the 
interacting
system corresponds exactly to one of the non-interacting system with the same quantum number 
$\{I_j\}$ and each allowed
$I_j$ may describe a quasiparticle state. Without external perturbation, the distribution 
$\{I_j\}$
is time independent. Therefore, in a pure quantum state we can say a quasiparticle has an infinite life-time 
and is thus well defined.
In the ground state, only $N$ (particle number) states of $I_j$ with the 
smallest absolute values are occupied. These occupied $I_j$ states form a 
perfect Fermi sphere. In such a sense, the Luttinger theorem is satisfied. In
 addition, the particle-hole excitations can be constructed by moving some 
particles from the Fermi sphere (thus leaving some $I_j$ holes) up to higher
 $I_j$ states. This can be exactly shown in the integrable models\cite{10}.
 In these models, the excitation energy is exactly given by $\Delta 
E=\sum_p\epsilon(I_p)-\sum_h\epsilon(I_h)$, where $\epsilon(I_p)$ 
($\epsilon(I_h)$) is the dressed energy of the quasiparticles 
(holes)\cite{10}. Also, the exact creation (annihilation) operator of the
 quasiparticles $R^\dagger(p)$ ($R(p)$) for the integrable models can be 
constructed in the framework of the algebraic Bethe ansatz\cite{11}. Though 
these operators satisfy an nonlocal commutation relation 
$R^\dagger(p)R^\dagger(q)=S(p,q)R^\dagger(q)R^\dagger(p)$, the Pauli exclusion principle
 holds for these quasiparticles since the scattering matrix 
$S(p,q)$ takes the value -1 when $p=q$. In fact, $R^\dagger(p)$ is the exact creation operator 
of the eigenstates and has the perfect time evolution form 
$R^\dagger(p,t)=exp(i\epsilon(p))R^\dagger(p,0)$. 
\par
Based on the above discussions, we can see that at least for the equilibrium state,
Landau's Fermi liquid theory is applicable in 1D. Caution should be taken only for
the transport properties and other non-equilibrium properties. We note a different approach, the so-called
Landau-Luttinger-liquid theory\cite{12} based on the exact solution of the Hubbard model was established
by Carmelo et al.. The quasi-particles are defined in the charge- and spin-sectors respectively rather than in
the conventional way. This is the poineering work approaching to the 1D quantum systems along the line
of the generalized Fermi-liquid theory and obtained the related quantities in a microscopic way\cite{13,14}.
 The main conclusion of the present letter is that almost
all the known results derived from the LLT can be retrieved from 
Landau's phenomenological FLT. The one-to-one correspondence between the Landau parameters and the 
Haldane parameters can be established. The present results strongly suggests that the Luttinger-liquid, 
the Landau-Luttinger liquid and  Landau's conventional Fermi liquid are essentially paralell in 1D.
\par
A microscopic state of the Fermi liquid is  described by 
the quasiparticle distribution $\{n_{ps}\}$.  In an equilibrium state, $n_{ps}$ takes the well 
known Fermi-Dirac form $n_{ps}=\{exp[(\epsilon_p-\mu)/T]+1\}^{-1}$, 
where $\mu$ is the 
chemical potential. Notice that the quasiparticle energy $\epsilon_p$ is no longer a c-number but a 
functional of $\{n_{p's'}\}$, 
\begin{eqnarray}
\epsilon_{p}=\epsilon_{p}^0+\frac 1L\sum_{p's'}f_{ps,p's'}\delta 
n_{p's'}+\cdots,
\end{eqnarray}
where $\epsilon_{p}^0$ is the energy of the quasiparticles in the ground 
state, and $f_{ps,p's'}$ is the Landau function which describes the 
interaction of the quasiparticles.
Since we consider the low energy properties of the system, only the events
 near the Fermi surface are important. Therefore, we shall replace 
$f_{ps,p's'}$ by its values at the Fermi points $p=\pm p_F, p'=\pm p_F$. The 
Fermi momentum is given by $p_F=\pi n/2$. For convenience, we introduce the 
following notations:
\begin{eqnarray}
f_0^s=\frac12\sum_{r=\pm}\sum_{s'}f_{p_Fs,rp_Fs'},{~~}f_1^s=\frac12\sum_{r
=\pm}\sum_{s'}rf_{p_Fs,rp_Fs'},\nonumber\\
f_0^a=\sum_{r=\pm}\sum_{s'}s'f_{p_Fs,rp_Fs'},{~~}f_1^a=\sum_{r=\pm}\sum_{s
'}rs'f_{p_Fs,rp_Fs'},\\
F_{0,1}^{s,a}=N(0)f_{0,1}^{s,a},{~~~~~~~~~~~~~~~~~~~}\nonumber
\end{eqnarray}
where $N(0)$ is the density of states at the Fermi surface.
\par
As the momentum per unit length is just the mass flux of the system, we can easily
deduce the effective mass of the quasiparticles near the Fermi surface as
\begin{eqnarray}
m^*=\frac{p_F}{v_F}=1+F_1^s.
\end{eqnarray}
In another hand, the density of states per
 unit length of the quasiparticles at the Fermi surface is
\begin{eqnarray}
N(0)=\frac1L\sum_p\delta(\epsilon_p^0-\mu)=
\frac 1{\pi v_F},
\end{eqnarray}
where $v_F$ is the Fermi velocity. Notice that we have taken the particle mass $m$ as our unit.
To derive some important quantities such as the compressibility $\kappa$ and the Drude weight
$D$, we consider $\delta n_{ps}$ induced by the 
external pressure and flux $\Phi$
\begin{eqnarray}
\delta n_{ps}=\frac{\partial n_{ps}}{\partial \epsilon_p}(\delta 
\epsilon_p-\delta\mu-\frac p{|p|}\delta \Phi).
\end{eqnarray}
At low temperatures, with eq.(3) we find the following relations hold
\begin{eqnarray}
\frac{\partial n}{\partial \mu}=\frac{2N(0)}{1+F_0^s},{~~~~~}\frac 1L\frac{\partial J}{\partial \Phi}
=\frac{2N(0)}{1+F_1^s},
\end{eqnarray}
where $J=\sum_{p,s}p/|p|n_{ps}$ is the chiral current. From this we obtain
\begin{eqnarray}
\kappa\equiv\frac 2{\pi n^2C_{N_\rho}}=\frac 2{\pi n^2v_F}\frac 1{1+F_0^s},\nonumber\\
D\equiv\frac 2{\pi C_{J_\rho}}=\frac2{\pi
 v_F}\frac1{1+F_1^s}.
\end{eqnarray}
Similarly, the following relations hold
\begin{eqnarray}
\frac{\partial M}{\partial H}=\frac{2N(0)}{1+F_0^a},{~~~~}\frac1L\frac{\partial J_\sigma}{\partial \Phi_\sigma}=\frac{2N(0)}
{1+F_1^a},
\end{eqnarray}
and the magnetic susceptibility $\chi$ and the spin Drude weight $D_\sigma$ are given by
\begin{eqnarray}
\chi\equiv\frac 1{\pi C_{N_\sigma}}=\frac 1{\pi v_F}\frac 1{1+F_0^a},\nonumber\\
D_\sigma\equiv\frac 1{\pi C_{J_\sigma}}=\frac1{\pi 
v_F}\frac1{1+F_1^a},
\end{eqnarray}
where $M$ and $J_\sigma$ are the magnetization and the spin current induced by 
the magnetic field $H$ and the spin flux $\Phi_\sigma$ respectively.
Notice above four new parameters $C_{N_\rho}, C_{J_\rho}$, $C_{N_\sigma}, C_{J_\sigma}$ are defined. They are nothing
but the zero sounds in a 1D Fermi liquid. As the temperature is lowered, the mean quasiparticle 
scattering time $\tau$ increases and, for fixed frequency, $\omega \tau$ 
increases. As $\omega\tau$ nears one, the quasiparticles no longer have time
 to relax in one period of the sound; the liquid then no longer remains in 
local thermodynamic equilibrium, and the character of the sound propagation 
begins to change. If there is no quasiparticle collision, $f_{ps,p's'}$, then
  in the collision regime $\omega\tau >>1$, the excess quasiparticles in a 
fluid element with increased density simply diffuse away, without driving 
the neighboring elements. However, as Landau observed, if there are 
quasiparticle interactions, then a local increase in the density of a fluid 
element can drive neighboring elements via the modification of the effective
 field $\sum_{p's'}f_{ps,p's'}\delta n_{p's'}(x,t)$. Such restoring forces 
between neighboring elements can give rise to a sound-like collective modes 
of oscillation of the fluid, called zero sound.
\par
For the liquid in nonequilibrium and inhomogeneous situations that differ 
slightly from the equilibrium state of the homogeneous liquid, the state of 
the liquid can be specified by the quasiparticle distribution $n_{ps}(x,t)$ 
as a function of position and time, which satisfies the Landau kinetic 
equation\cite{1,9}
\begin{eqnarray}
\frac{\partial n_{ps}(x,t)}{\partial t}+\frac {\partial}{\partial
 p}\epsilon_p(x,t)\frac{\partial}{\partial 
x}n_{ps}(x,t)\nonumber\\
-\frac{\partial}{\partial x}\epsilon_p(x,t)\frac 
{\partial}{\partial p}n_{ps}(x,t)=I[n_{p's'}],
\end{eqnarray}
where $I[n_{p's'}]$ is the collision integral. We remark that the above equation may not be reliable in 1D due to the
short collision time. However, for
$p=\pm p_F$ it is correct in any dimensions, since in this case the quasi-particles have an infinite long life-time, 
 and the collision integral is exactly zero. In this case, (12) can be linearized approximately as
\begin{eqnarray}
(\frac{\partial}{\partial t}+v_p\frac{\partial}{\partial x})\delta 
n_{ps}(x,t)-(\frac{\partial n_{ps}^0}{\partial 
\epsilon_p})v_p\frac{\partial}{\partial x}\delta \epsilon_p(x,t)=0,
\end{eqnarray}
where $\delta\epsilon_p(x,t)$, the effective field that drives the 
quasiparticles, is given by
\begin{eqnarray}
\delta\epsilon_p(x,t)=U(x,t)+\frac 1L\sum_{p's'}f_{ps,p's'}\delta 
n_{p's'}(x,t).
\end{eqnarray}
Assuming the external field $U(x,t)= U\exp{i(qx-\omega t)}$, we have 
$\delta n_{ps}(x,t)=\delta n_{ps}(q,\omega)\exp{i(qx-\omega t)}$, and
\begin{eqnarray}
\frac{\partial n_{ps}^0}{\partial 
\epsilon_p}qv_p(U+\frac 1L\sum_{p's'}f_{psp's'}\delta n_{ps}(q,\omega))\nonumber\\
+(\omega-qv_p)\delta n_{ps}(q,\omega)=0.
\end{eqnarray}
If we write $\delta n_{ps}$ in the form $\delta n_{ps}=-\partial 
n_{ps}^0/\partial \epsilon_p \nu_{ps}$, then the $\nu_{ps}$ obey
\begin{eqnarray}
\nu_{ps}-\frac{qv_p}{\omega-qv_p}N(0)\sum_{p's'}f_{ps,p's'}\nu_{p's'}=
\frac{qv_p}{\omega-qv_p}U.
\end{eqnarray}
Notice above $p=\pm p_F$ and $v_{\pm p_F}=\pm v_F$. By solving the
 four coupled linear equations, we easily obtain the velocities of the four
 branch zero sounds
\begin{eqnarray}
C_\rho^0=v_F(1+F_0^s), {~~~~}C_{J_\rho}^0=v_F(1+F_1^s),\nonumber\\
C_\sigma^0=v_F(1+F_0^a), {~~~~}C_{J_\sigma}^0=v_F(1+F_1^a).
\end{eqnarray}
These quantities are exactly the same given in (9) and (11). The former two correspond to the zero sounds of charges  and the latter
 two correspond to those of the spins, respectively. In fact, the zero sounds describe the motions of the whole
 Fermi sphere's deformation in the extreme collisionless
limit $\omega\tau >>1$. For each sector (spin and charge), there are only two
 such modes in 1D. One is the breath mode of the Fermi sphere  and the other is 
the oscillation of the whole Fermi sphere around the equilibrium position. 
The situation is very different from that of the Fermi liquid in 3D, where 
there are an infinite number of zero sounds. 
\par
At frequencies $\omega$ sufficiently small that 
$\omega \tau<<1$, 
sound in Fermi liquids takes the form of ordinary hydrodynamic; or first 
sound. The sound velocity $C_\rho$ can be given by
\begin{eqnarray}
C_\rho=\sqrt{n\partial \mu/\partial n}=v_F\sqrt{(1+F_0^s)(1+F_1^s)}.
\end{eqnarray}
In fact, in the hydrodynamic limit, the density fluctuation $\delta n(x,t)$ satisfy the
following equation
\begin{eqnarray}
\frac{\partial^2\delta n}{\partial t^2}-\frac{\partial^2}{\partial
 x^2}\delta P=0,
\end{eqnarray}
 where $\delta P$ is
 the variation of the pressure and can be 
approximated by $n(\partial \mu/\partial n)_{T=0}\delta 
n={C_\rho}^2\delta n$. If we put $\delta n(x,t)$ as a quantum field rather than a semiclassical quantity, with the
canonical quantization
\begin{eqnarray}
[\delta n(x,t),\partial_t\delta n(x',t)]=i\delta(x-x'),
\end{eqnarray}
the density fluctuations in 1D Fermi liquid can be simply ``bosonized" as
\begin{eqnarray}
H_b=\int[(\frac{\partial\delta n}{\partial t})^2+{C_\rho}^2(\frac{\partial\delta n}{\partial x})^2]dx.
\end{eqnarray}
Such an  effective Hamiltonian is 
exactly the same derived in the Luttinger liquid theory. The spin density fluctuations can be quantized in a similar way with
the spin wave sound velocity
\begin{eqnarray}
C_\sigma=v_F\sqrt{(1+F_0^a)(1+F_1^a)}.
\end{eqnarray}
In 1D, the sound
 waves can be understood as the linear combination of an infinite number of 
quasiparticle-hole excitations with the same momentum $q$. In such a sense,
 the low-lying quasiparticle-hole excitations can be completely included in
 the collective modes, i.e., the first sounds. The specific heat of the 
system is thus given by
\begin{eqnarray}
C=\frac \pi 3(\frac 1{C_\rho}+\frac 1{C_\sigma})T.
\end{eqnarray}

\par
By comparing the measurable quantities derived from FLT and
 those from LLT, we can easily deduce that an exactly one-to-one
 correspondence exists between the parameters of the 1D Fermi liquid and 
those of the Luttinger liquid:
\begin{eqnarray}
C_{N_\gamma}=v_{N_\gamma},{~~~~}C_{J_\gamma}=v_{J_\gamma},{~~~~}C_\gamma=
v_\gamma,{~~~~~}\gamma=\sigma,\rho.
\end{eqnarray}
In addition, the Haldane relations are also satisfied in the Fermi liquid
 description
\begin{eqnarray}
C_{N_\gamma}C_{J_\gamma}=C_\gamma^2, {~~~~~}C_{J_\gamma}=K_\gamma C_\gamma,
\end{eqnarray}
where the stiffness constants in the Luttinger liquid theory can be expressed
 by the Fermi liquid parameters as
\begin{eqnarray}
K_{\rho,\sigma}=\sqrt{\frac{1+F_1^{s,a}}{1+F_0^{s,a}}}.
\end{eqnarray}
 An interesting fact
 is that all the quantities of the forward scattering g-ology model (spin-1/2
 Luttinger model) obtained from the bosonization method\cite{5,6} are exactly
 the same to those derived from FLT with the notation 
correspondence: $g_{ss'}^4=f_{p_Fs,p_Fs'}, g_{ss'}^2=f_{p_Fs,-p_Fs'}$. Such a 
correspondence has also been pointed out in a recent review\cite{15}. 
 
\par
 The correlation functions can not be derived from FLT in a microscopic way since the theory itself is
phenomenological. However, we know that 1D quantum systems are conformally invariant at 
zero temperature\cite{16} and the exponents of a variety of correlation functions
are directly related to the finite size corrections of the corresponding
elementary excitations\cite{17} which can be easily derived from the present theory. 
In a 1D Fermi liquid, there are only three types
of elementary excitations in each sector, i.e., the small momentum particle-holes,
the $2k_F$ particle-holes or the currents and the additional charges. The finite size 
correction for the excitation
energy of a 1D Fermi liquid can be derived from (8), (10) and (21)as
\begin{eqnarray}
\Delta F=\sum_{\gamma=\rho,\sigma}\frac{2\pi C_\gamma}L[I_\gamma^++I_\gamma^-+
\frac{{\Delta N_\gamma}^2}{8K_\gamma}+\frac18K_\gamma{\Delta J_\gamma}^2],
\end{eqnarray}
where $I_\gamma^{\pm}$ are integers and the superscripts $\pm$ character the two Fermi points, $\Delta N_\gamma$
and $\Delta J_\gamma$ are the numbers of additional charges and currents respectively.
What a nice thing is that all the critical exponents we obtained are just the same to those derived
from LLT\cite{6}.
\par
Now we turn to the impurity problem in 1D. As pointed out by Kane and Fisher\cite{18}, an
 impurity in a 1D gapless system cuts the chain into disconnected and therefore 
has a serious effect to the transport properties, while those in 3D only induce the residual
resistivity. Generally,
a $(d-1)$-dimensional defect in a $d$-dimensional host must has a remarkable effect to
the transport coefficients in the direction perpendicular to the defect ``plane".
A point impurity in 1D falls the case while in 3D does not. The transport of electrons
through the impurity in 1D can be described as a tunneling effect\cite{19} and is 
perturbationable based on the open boundary fixed point. Such a system is equivalent to a periodic
one with doubled length and doubled particle number. However, the quasiparticle states
$(k,\sigma), (-k,\sigma)$ correspond to the same state due to the reflection symmetry.
That means the current excitation is forbidden. With the predictions of the boundary conformal
field theory\cite{20}, the anomaly dimensions of a variety boundary operators can be derived
from the finite size correction as
\begin{eqnarray}
x_b=\sum_{\gamma=\rho,\sigma}[I_\gamma+\frac{{\Delta N_\gamma}^2}{4K_\gamma}],
\end{eqnarray}
where $I_\gamma$ are non-negative integers characterizing the particle-hole excitations. Suppose the
impurity is sited at the origin. The tunneling current through it can be written as
\begin{eqnarray}
J(x,y)\sim -i[C_\sigma^\dagger(x)C_\sigma(y)-h.c],{~~~}x\sim 0^+,{~~}y\sim 0^-.
\end{eqnarray}
>From (28) we obtain the asymptotic time correlation function
\begin{eqnarray}
<[J(x,y|\tau), J(x,y|0)]>\sim \tau^{-K_\rho^{-1}-K_\sigma^{-1}}.
\end{eqnarray}
Therefore, the tunneling resistivity for the spin-rotational invariant systems ($K_\sigma=1$) has the asymptotic form
\begin{eqnarray}
\rho_i(T)\sim T^{1-K_\rho^{-1}},
\end{eqnarray}
which also coincides with the result derived from LLT\cite{18,21,22}.
\par
The author acknowledges the financial supports of China National Foundation of
Natural Science and Alexander von Humboldt-Stiftung.

\end{document}